\documentclass[useAMS,usenatbib]{mn2e}

\usepackage[psamsfonts]{amssymb}
\usepackage[dvips]{graphicx}
\usepackage{amsmath,alltt}
\usepackage{multirow}
\usepackage{rotating}
\usepackage{lscape}
\title[The Dynamical Significance of Triples]{The Dynamical Significance of Triple Star Systems in Star Clusters}
\author[Leigh \& Geller]{Nathan W. C. Leigh$^{1}$, Aaron M.~Geller$^{2}$
\thanks{E-mail: nleigh@rssd.esa.int (NL); a-geller@northwestern.edu (AG)}\\
$^{1}$European Space Agency, Space Science Department, Keplerlaan 1,
2200 AG Noordwijk, The Netherlands \\
$^{2}$Center for Interdisciplinary Exploration and Research in Astrophysics (CIERA) \&  Dept. of Physics and Astronomy, \\ Northwestern University, 2145 Sheridan Rd, Evanston, IL 60208, USA}
\begin{document}

\pagerange{\pageref{firstpage}--\pageref{lastpage}} \pubyear{2011}

\maketitle

\label{firstpage}

\begin{abstract}
Over the last few decades, observational surveys have revealed that 
high-order multiple-star systems (e.g. triples, quadruples, etc.), and triples in particular, 
are common in our Galaxy.  
In this paper, we consider the dynamical significance of this transformation in 
our understanding of stellar multiplicity.  Using empirically constrained 
binary and triple fractions in those star clusters for which these values are available 
in the literature, 
we compare analytic rates for encounters involving single, binary, and triple stars.  
Our results show that, \textit{
even for relatively low triple fractions,
dynamical interactions involving triples 
occur roughly as often as encounters involving either single or binary stars 
alone, particularly in low-mass star clusters}.  More generally, using empirically-derived multiple 
star catalogues for the young star-forming association Taurus-Auriga and
the Galactic field, we show that the data are consistent with the 
gravitationally-focused cross section for encounters increasing 
with increasing multiplicity.  Consequently, triple stars, and even higher-order 
multiples, could be more important than previously realized for a number of 
astrophysical phenomena, including 
the formation and destruction of compact binaries and various types of
stellar exotica, and the dynamical evolution of star clusters.
%, and the production
%of high-velocity escapers during resonant encounters.

\end{abstract}

\begin{keywords}
gravitation -- stellar dynamics -- binaries (including multiple): close -- stars: formation 
-- globular clusters: general -- open clusters and associations: general.
\end{keywords}

\section{Introduction} \label{intro}

%AMG: have you read this paper : http://adsabs.harvard.edu/abs/2012AJ....144...62A
%I think this should be included somewhere in the introduction, and maybe we can use something from their Table 1 in our analysis?
%NL: We now discuss this paper in the intro, a few paragraphs down.  I can't see any way to integrate table 1 into things though...
%AMG: great addition to the intro.  I guess I was thinking to use the info in their Tables 1 and 2 to get the a_b/a_t value or maybe more points on Figure 3?  But it's your call whether you think that work is worth it.  (We may already have our hands full explaining the observed samples we have in here now :)

Over the last few decades, observations have revealed that multiple-star systems 
(MSSs), including 
binary stars and high-order multiples (e.g, triples, quadruples, etc.),
are common in our Galaxy.  This is the case for 
the Galactic field \citep[e.g.][]{duquennoy91,tokovinin97,raghavan10}, young 
star-forming associations \citep[e.g.][]{kraus11,kraus12}, old moderately-dense 
open clusters \citep[e.g.][]{latham07,talamantes10,geller12a}, and even ancient 
globular clusters (GCs) \citep[e.g.][]{prodan12,milone12}.  
The most complete survey is that of \citet{raghavan10}, who 
recently updated the seminal work of \citet{duquennoy91} using a 
volume-limited sample of field solar-type primary stars in the solar 
neighborhood selected from the \textit{Hipparcos} catalogue.  They reported 
that the observed fractions of objects that are single, double, triple, and high-order 
systems are $56 \pm 2\%$, $33 \pm 2\%$, $8 \pm 1\%$, and $3 \pm 1\%$, 
respectively.  
Even more recently, \citet{kraus11} 
performed a high-resolution imaging study of the young star-forming 
region Taurus-Auriga to characterize its multiple star population.  They 
found that $\sim 2/3 - 3/4$ of all targets are MSSs 
composed of two or more stars, while 
only the remaining $\sim 1/4 - 1/3$ are single stars. 

The importance of dynamical encounters involving MSSs, 
even during the epoch of star formation, is becoming increasingly evident.  
\citet{bate12} recently presented the results of a statistical analysis of 
MSSs formed during the largest radiation hydrodynamical 
simulation of star cluster formation conducted to date.  
These models reproduce reasonably well many 
of the characteristics of Galactic star-forming regions.  
%, including
%multiplicity as a function 
%of primary mass, the frequency of very low-mass binaries, and correlations 
%involving the mass ratio and separation distributions of binaries.  
Interestingly, the author finds that
protostellar accretion is often terminated by dynamical interactions,
and therefore that dynamics plays an important role in shaping the
stellar initial mass function (IMF) \citep{podsiadlowski92}, at least in 
the dense clusters considered in this study.  The importance of dynamics and 
multiplicity during the star formation process has also been confirmed 
observationally.  Based on a multi-epoch search 
for wide, low-mass tertiary companions in a volume-limited sample of 
118 known spectroscopic binaries
%AMG: added
within 30 pc of the Sun, \citet{allen12} find a wide tertiary fraction 
of $19.5^{+5.2}_{-3.7}$\%.  This is consistent with the predictions of star 
formation simulations, which suggest that the fraction of wide, low-mass 
companions to spectroscopic binaries is $>10$\%, and roughly twice the 
wide companion rate of single stars.  This trend is thought to arise through 
three-body interactions during the star formation process, which transfers 
angular momentum away from a close pair of objects, hardening them 
further.  

Our theoretical understanding of how dynamical encounters involving high-order multiples 
impact the evolution of older star clusters is 
limited \citep[e.g.][]{moeckel13}.  The vast majority of the 
numerical simulations of star cluster evolution performed to date include 
only single stars and binaries in the initial population, and 
often these models do not allow for the dynamical formation of 
high-order multiples, 
thus entirely neglecting the dynamical input from this important stellar population.
Some authors have recently 
acknowledged the need to include high-order multiples in their 
simulations, and list this as a future addition 
\citep[e.g.][]{hypki12}. 
\citet{geller13} investigate the dynamical impact of both primordial and 
dynamically formed triples on blue straggler formation in their $N$-body models, and 
find that nearly half of the blue stragglers formed through collisions (between 
6 and 7.5 Gyr) resulted from stellar encounters involving hierarchical triples.
We suggest here that the inclusion of high-order multiples in such models 
may be critical for accurately reproducing the dynamical evolution of many star
clusters as well as the production rates of compact binaries and stellar exotica.

%In this Letter, we address the question:  
%How important are high-order MSSs to the dynamics in 
%star clusters?  This remains largely 
%an open question in astrophysics.  In an attempt to guide us toward a 
%preliminary answer, we present 
%three principles of multiplicity (Section~\ref{principles})
%applicable to dynamical interactions in open and globular star 
%clusters.  These are meant to provide 
%a clear, concise application of previous observational 
%and theoretical results to dynamical interactions involving MSSs.  Their 
Recently, \citet{leigh11} argued that encounters involving triple stars 
should be common in the old open clusters NGC 188 and M67, and that this 
could also be the case for other open clusters.  
Importantly, however, these results 
were based on the \textit{inferred} properties of the binary and triple 
populations in these clusters.  In this paper, we put the prediction 
of \citet{leigh11} to the test using \textit{empirically constrained} binary 
and triple fractions in those open clusters for which these data are 
available in the literature.  We show in Section~\ref{dynsig} that 
encounters involving triples occur roughly as often as, or even more often 
than, encounters 
involving single and binary stars 
%AMG: added ``alone'' (again)
alone in every cluster in our sample.  
More generally, using empirically-derived multiple star
catalogues for the young star-forming association Taurus-Auriga and
the Galactic field, we show in Section~\ref{multiples} that the 
gravitationally-focused cross section for encounters increases 
with increasing multiplicity.  This reduces the average time between 
encounters and, in general, contributes to increasing the dynamical 
significance of high-order multiple star systems.  
The implications of our results to various astrophysical phenomena 
are discussed in Section~\ref{discussion}, and we summarize the 
key points of this paper in Section~\ref{summary}.

\section{Identifying the Dominant Encounter Type} \label{dynsig}

In this section, we compare the rates of the different encounter types involving 
singles, binaries, and triples using empirically-measured binary and triple fractions 
taken from the literature.  

\begin{figure}
\begin{center}
\includegraphics[width=\columnwidth]{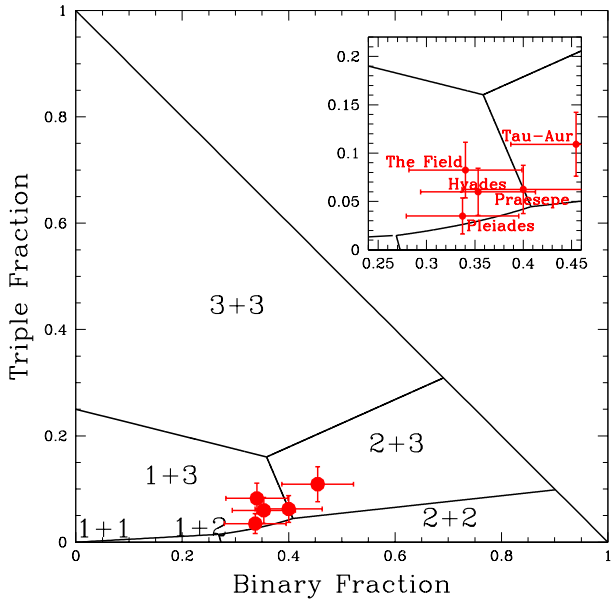}
\end{center}
\caption[The relative rates of the different encounter types in the
binary fraction - triple fraction plane for Taurus-Auriga,
Praesepe, the Pleiades, the Hyades and the field]{The relative rates of the
different encounter types in the
  binary fraction - triple fraction plane for Taurus-Auriga, Praesepe,
the Pleiades, the Hyades, and
the Galactic field, as shown by the red points.  The ratios between
the geometric
cross-sections for single, binary, and triple stars observed for
Taurus-Auriga were used to calculate the locations of the black lines
that divide the parameter space in the binary fraction - triple fraction
plane for which each of the different encounter types dominates.
Error bars on the observations denote the $1\sigma$ uncertainties calculated using 
Poisson statistics.
\label{fig:fb-ft}}
\end{figure}

\citet{leigh11} derive encounter rates for single-single (1+1), single-binary (1+2), 
binary-binary (2+2), single-triple (1+3), binary-triple (2+3), and triple-triple (3+3) 
encounters.  Importantly, the ratio of any two of these encounter rates
only depends on the binary and triple fractions, and the average geometric cross sections for
singles, binaries and triples (e.g., the mean stellar radius, $\bar{R}$, and the mean binary
and tertiary semi-major axes, $\bar{a}_b$ and $\bar{a}_t$).
Therefore, the regions of parameter space where each of these different encounter rates dominate
can easily be compared across these three different stellar populations, as
we do in Figure~\ref{fig:fb-ft}.  Each labeled polygon in this figure denotes the region
of parameter space where the frequency of the given encounter type is higher than all others,
and dividing lines show where the encounter frequencies are equal.  Here we show the result
using the value of $\bar{a}_t/\bar{a}_b$ for Taurus-Auriga\footnote{Specifically,
$\bar{a}_t$ = 1773.7 AU and $\bar{a}_b$ = 130.0 AU, and assuming
$\bar{R}$ = 1.5 R$_{\odot}$.}.
Perhaps the most striking result from this analysis is that
encounters involving triples dominate over encounters with binaries for the
majority of parameter space.  This is the case on a per star basis, even if the fraction of
systems composed of 3 or more stars is low.\footnote{We define the binary and triple fractions
as $f_b = N_b/(N_s+N_b+N_t)$ and $f_t = N_t/(N_s+N_b+N_t)$, respectively, where $N_s$, $N_b$,
and $N_t$ are the numbers of
single, binary, and triple stars, respectively (ignoring multiplicities higher than three in
these definitions).}

\begin{figure}
\begin{center}
\includegraphics[width=\columnwidth]{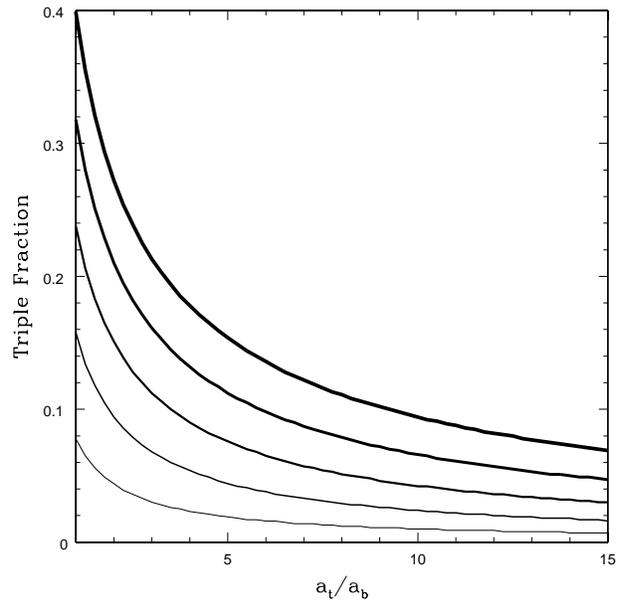}
\end{center}
\caption[For constant binary fraction, the critical triple fraction at which encounters 
involving triples begin to dominate]{Plot showing, for lines of constant binary fraction, 
the critical triple fraction at which encounters involving triples begin to dominate over 
encounters involving singles and binaries alone as a 
function of the ratio $\bar{a}_t/\bar{a}_b$.  
%Black and red lines separate the parameter space where encounters involving triples 
%dominate over encounters involving binaries and single stars, respectively.  
Different binary fractions are shown using different line widths.  The line 
width increases with increasing binary fraction, and the results are shown for 
$f_b = 0.1, 0.2, 0.3, 0.4, 0.5$.
\label{fig:fig2}}
\end{figure}

%AMG: see my email about modifying this figure and also the text here slightly.
%NL: Modified on all accounts.
In Figure~\ref{fig:fig2} we show, for lines of constant binary fraction, 
the critical triple fraction at which encounters involving triples begin to dominate
over encounters involving single and binary stars alone.  To obtain each 
of these lines, we fix $f_b$ and calculate the number of encounters involving 
triples (i.e. the sum of the numbers of 1+3, 2+3, and 3+3 encounters) over a fixed 
time-scale.  
For the same time-scale, we then compare this to the 
total number of encounters involving binaries and singles alone (i.e. 1+1, 1+2, and 
2+2).  
The critical triple fraction is shown as a function of the ratio $\bar{a}_t/\bar{a}_b$. 
Importantly, if $f_t \gtrsim 0.3$, encounters 
involving triples will dominate over encounters involving either single or
binary stars, regardless of $f_b$ and the ratio $\bar{a}_t/\bar{a}_b$.  
Equivalently, provided $\bar{a}_t/\bar{a}_b \gtrsim 5$ and $f_t \gtrsim 0.16$, 
triples will be undergoing encounters more often than either single or binary 
stars
%AMG: added alone
alone.  Furthermore, for $f_t \gtrsim 0.1$, if triples are not the dominant type of 
interacting object, they are at least 
roughly as dynamically-active as are single and/or binary stars.

%Recall that Figure~\ref{fig:fb-ft} is constructed using the Taurus-Auriga 
%$\bar{a}_t/\bar{a}_b$ value, which 
%may not be the same for all clusters shown here.  Nevertheless, encounters involving 
%triples remain comparably frequent to those involving either binary or single stars 
%provided $\bar{a}_t > \bar{a}_b$.  This is demonstrated 
%in Figure~\ref{fig:fig2}, which shows, for lines of constant binary fraction, 
%the critical triple fraction at which encounters involving triples begin to dominate, 
%either over binaries (black lines) or single stars (red lines).  The critical 
%triple fraction is shown as a function of the ratio $\bar{a}_t/\bar{a}_b$.  A few 
%things are apparent in Figure~\ref{fig:fig2}.  
%First, for $f_b \lesssim 0.10$, 
%triples will dominate over singles only if $f_t \gtrsim 0.17$, almost independent 
%of the ratio $\bar{a}_t/\bar{a}_b$.  This implies that encounters involving 
%singles will nearly always dominate over encounters involving triples 
%provided $f_b \lesssim 0.15$, since otherwise $f_t > f_b$, which is inconsistent 
%with every MSS sample known in the literature.  Second, if $f_t \gtrsim 0.3$, encounters 
%involving triples will always dominate over encounters involving both single and 
%binary stars, and this is independent of $f_b$ and the ratio $\bar{a}_t/\bar{a}_b$.  
%More realistically, provided $\bar{a}_t/\bar{a}_b \gtrsim 5$ and $f_t \gtrsim 0.16$ 
%are both satisfied, encounters involving triples will dominate over those involving 
%both single and binary stars.

\begin{table*}
\caption{The numbers of single, binary, and triple stars in our samples.}
\begin{tabular}{|c|c|c|c|c|}
\hline
Cluster Name      &  Singles   &   Binaries   &   Triples  &  Total   \\
\hline
Taurus-Auriga     &     48     &      50      &     12     &   110    \\
Galactic Field    &     56     &      33      &      8     &    97    \\
Hyades            &     98     &      59      &     10     &   167    \\
%Praesepe          &     47     &      30      &      3     &    80    \\
Praesepe          &     43     &      32      &      5     &    80    \\
%Pleiades          &     56     &      30      &      2     &    88    \\
Pleiades          &     54     &      29      &      3     &    86$^a$    \\
%AARON:  Could you please update the below? - all I have are the binary/triple fractions from you, not total numbers.  Thanks.
%AMG: I think we should just remove NGC 188.  See below
%NGC 188           &     71     &      28      &      1     &   100    \\
\hline
%\end{tabular}
\multicolumn{5}{c}{$^a$The original \citet{mermilliod92} sample contains 88 stars, but four of these stars are associated as visual binaries, respectively.} \\
\end{tabular}  
\label{table:stats}
\end{table*}

Next we compare these analytic results to observed multiple-star populations from the literature. 
The points in Figure~\ref{fig:fb-ft} show the observed binary and triple fractions for
Taurus-Auriga \citep{kraus11}, the Pleiades \citep{mermilliod92, bouvier97}, Praesepe \citep{mermilliod99,bouvier01},
and the Hyades \citep{patience98}.
%, and NGC 188 \citep{geller08,geller09,geller12a}.
We also place a point on Figure~\ref{fig:fb-ft} representing a star
cluster with the binary and triple fractions observed in the Galactic field \citep{raghavan10}.  
We plot these empirical results for illustrative purposes, but note that each survey has a unique 
level of completeness and sample selection criteria, which we do not attempt to correct for here.  
The raw numbers of single, binary, and triple stars are also shown for each surveyed cluster in
Table~\ref{table:stats}.  Below, we briefly describe the individual samples.

The most complete surveys shown here are likely those of the Galactic field, Taurus-Auriga, and the Hyades.  
The \citet{raghavan10} Galactic field survey covers a complete sample of solar-type dwarfs within 25 pc of the Sun
having $0.5 \leq (B-V) \leq 1.0$ (which roughly corresponds to spectral types F6 - K3). 
This study combines multiple stars detected through spectroscopy, eclipsing binary surveys, separated fringe packets 
(using the Center for High Angular Resolution Astronomy, CHARA, Array), speckle interferometry, adaptive optics, and 
multiple astrometric techniques (e.g., astrometric orbits, common-proper motion pairs,  etc.).
This study is nearly complete for orbital separations $\lesssim 10^4$ AU and mass ratios $\gtrsim 0.1$.
The \citet{kraus11} study contains all known members of Taurus-Auriga with spectral types between G0 and M4
and masses between 2.5 M$_{\odot}$ and 0.25 M$_{\odot}$.  Kraus et al.\ combine visual companions detected in 
2MASS images with companions detected through aperture masking with adaptive optics imagers to compile 
a fairly complete sample of multiple stars with separations between 3 and 5000 AU and mass ratios $\gtrsim 0.1$.
The \citet{patience98} Hyades speckle imaging survey covers
a nearly complete sample of stars with $K < 8.5$ mag (including dwarf stars of spectral type between about 
A0 and K5, and some evolved stars), stellar separations of about 5-50 AU and mass ratios $\gtrsim 0.2$.
The Hyades $f_b$ and $f_t$ values shown here also include multiple stars detected from 
spectroscopy and direct imaging (see references in \citealt{patience98}). 

The Pleiades and Praesepe samples are somewhat less complete.
For both clusters, we take the samples from the radial-velocity suverys of \citet{mermilliod92} and \citet{mermilliod99}, respectively,
and add additional companions detected in the adaptive optics imaging surveys of \citet{bouvier97} and \citet{bouvier01}, respectively.
For both clusters, the radial-velocity and adaptive optics samples do not completely overlap, and the coverage in moderate orbital separation 
is incomplete.  The \citet{mermilliod92} Pleiades survey covers stars with $0.4 < (B-V) < 0.9$  (about F5-K0 spectral types), while the 
\citet{bouvier97} survey includes stars with $0.56 \leq (B-V)_0 \leq 1.5$ (note, $E(B-V) = 0.03$) with known rotational velocities,
in the central part of the Pleiades.  
The combined Pleiades radial-velocity and adaptive optics numbers shown in Table~\ref{table:stats} are not sensitive 
to systems with separations between about 5 and 10-15 AU.
The \citet{mermilliod99} Praesepe survey covers stars with $0.4 < (B-V) < 0.8$ (about F5- K0 spectral types), while the 
\citet{bouvier01} survey includes proper-motion members with $0.52 < (B-V) < 1.4$.
The combined Praesepe radial-velocity and adaptive optics numbers shown in Table~\ref{table:stats} are not sensitive 
to systems with separations between about 4 and 15 AU.
For both clusters we also include photometric binaries identified by Mermilliod et al.\ as being in the binary region 
on the color-magnitude diagram (brighter and redder than the single-star main sequence), which may alleviate some of the incompleteness in 
orbital separation.

We note that the Pleiades, Praesepe and Hyades are all OCs
with ages on the order of several tens to hundreds of Myrs, and densities on the order 
of 1 - 10 M$_{\odot}$ pc$^{-3}$.  The ages of these clusters are 
comparable to the typical time between dynamical encounters, so that a large fraction 
of the MSSs in these clusters (and those in Taurus-Auriga) must be primordial.  

If these clusters have a similar ratio of $\bar{a}_t/\bar{a}_b$ as in Taurus-Auriga, 
Figure~\ref{fig:fb-ft} shows that all of these clusters
lie in the region where encounters involving triples will dominate over encounters involving both single and 
binary stars 
%AMG: added ``alone'' (again)
alone (not accounting for any incompleteness in the respective observational surveys).  
%AMG: added.  See if you like this, and change/add to it as you like.  Also, if you want to keep this, please add in the Bate numbers, and make sure that my text makes sense (I estimated a Bate ratio of ~1.5 from your figure).
%AMG: Also, can we get any at/ab values from the papers on the clusters (even though some are incomplete samples)?  Just a thought, could be useful, but might be more trouble than it's worth.
%For reference, the $\bar{a}_t/\bar{a}_b$ predicted by the \citet{bate12} model is xxx with a standard deviation of xxx.  
%AMG: moved this paragraph here, and made some edits

For reference, the \citet{geller13} $N$-body open cluster model predicts that, for dynamically formed solar-type triples in a rich open cluster
like NGC 188, the ratio of $\bar{a}_t/\bar{a}_b$ will remain roughly constant over the cluster lifetime (7 Gyr for their model), 
with a mean value of 4.2 and a standard deviation of 0.7.  
\citet{geller13} did not include any triples in the initial population.  Therefore their predictions are for triples formed 
during stellar encounters.  
%The discrepancy between these two theoretical predictions is intriguing, and may be a result, 
%in part, of the very different stellar densities of these simulations (with the \citealt{bate12} model significantly more dense),
%as well as a difference in triple formation with and without the presence of gas.  
In the Pleiades, the ratio $\bar{a}_t/\bar{a}_b$ is only a factor of $\sim 2-3$.  However the $1\sigma$ uncertainty exceeds 
this value and, as discussed 
%AMG: changed to above.
above,
%below, 
the Pleiades data are incomplete for moderate semi-major axes.
As also discussed above, the Taurus-Auriga data are the most complete of all of the open clusters in our sample, and
we note that the observed value in Taurus-Auriga of $\bar{a}_t/\bar{a}_b = 13.6$ is somewhat larger than the theoretical 
prediction of \citet{geller13} or the empirical value of the Pleiades.
%, but note that the 
%Taurus-Auriga observations are not sensitive to short-period ($< 3$ AU) systems. 
(We were unable to perform a similar comparison of the $\bar{a}_t/\bar{a}_b$ values for the other open clusters in our sample 
as these data were not readily available in the literature.)
%AMG: just to check, is the above statement correct?
We conclude that the available data are consistent with a typical ratio $\bar{a}_t/\bar{a}_b$ 
of at least a few, and possibly more.
Even if $a_b = a_t$, encounters involving triples occur roughly as frequently 
as encounters involving single stars, and encounters involving binaries occur 
at most $\sim 4$ times as often as those involving triples.  
%AMG: see comments in my email.  In short, I'm a bit confused with the comparison between encounters involving triples and those involving singles and/or binaries.  Of course your encounters involving triples also involve binaries and singles.  And do you include, say 1+3, in encounters involving single stars?  This is partly why I have added the word ``alone'' throughout, but make sure that this is correct .

Thus, we conclude that encounters involving triples should currently be 
as common as encounters involving only single or binary stars alone in every
cluster shown in Figure~\ref{fig:fb-ft} to within a factor of at most a few.  
In fact, the data are consistent with triples being the dominant interacting objects in these clusters.
Furthermore, given that in general triples are harder to detect than binaries 
(e.g., one may often require both spectroscopy and direct imaging for a given system to detect all three components)
the true locations of these clusters on Figure~\ref{fig:fb-ft} may lie even further into the 
triple dominated regime.  This can be properly tested when completeness-corrected period 
distributions become available for both binaries and triples in these clusters.  This will 
also allow for the results presented in this paper, which are calculated using \textit{average} 
cross-sections and hence \textit{average} encounter timescales, to be tested using timescales derived 
by integrating the full period and velocity distributions.

\section{Extension to Higher-Order Multiplicity} \label{multiples}

\begin{figure}
\begin{center}
\includegraphics[width=\columnwidth]{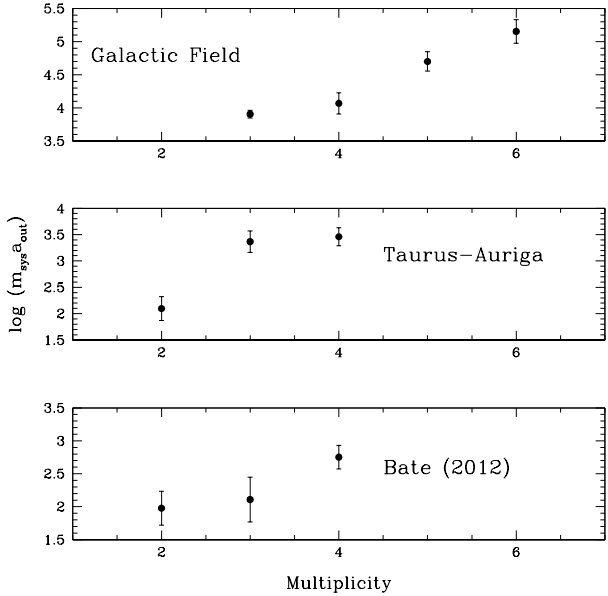}
\end{center}
\caption[The mean gravitationally-focused cross-section for encounters for 
all types of multiples in Taurus-Auriga, the Galactic Field, and the simulations 
of \citet{bate12}]{Plot showing the product of the mean total
system mass (m$_{\rm sys}$; in units of M$_{\odot}$) and the mean maximum orbital separation 
(a$_{\rm out}$; in units of AU), which is directly proportional 
to the mean or average gravitationally-focused cross-section for encounters, for
all types of multiples.  The top inset shows these cross-sections for the 
Galactic field, for which the results are shown for triples, quadruples, 
quintuples, and sextuples.  The middle inset shows the cross-sections for 
the young star-forming 
association Taurus-Auriga, and the results are shown for binaries, 
triples, and quadruples.  Finally, the bottom inset shows the cross-sections for the 
results of the hydrodynamical simulations of \citet{bate12}, for binaries, triples, and 
quadruples.  All error bars correspond to one standard 
%AMG: changed this like above, but be sure this is correct
error
%deviation 
of the mean.
\label{fig:fig3}}
\end{figure}

In this section, we extend our 
%AMG: edited
discussion
%results to include 
to higher-order multiples 
using the MSS samples of \citet{kraus11} and \citet{tokovinin97}.  These are 
representative of two different 
environments, namely a young (few Myrs) very low-density star-forming 
region and an older Galactic field population.\footnote{The field sample spans 
a wider range of spectral types, however our results remain unchanged if we 
limit this sample to cover the same spectral range as the Taurus-Auriga 
sample.  We therefore choose to use the complete field sample for Figure~\ref{fig:fig3}.}
%AMG: edited
%since the sample sizes are too small otherwise.}  
The MSC catalogue of \citet{tokovinin97} 
contains data on 612 (mostly) hierarchical multiple stars of multiplicity three to 
seven, collected both from direct imaging and spectroscopy.  The Taurus-Auriga 
MSS catalogue, on the other hand, consists of data from a high-resolution imaging 
study of 90 multiple star systems of multiplicity two to six with separations in 
the range 3-5000 AU.

The key point we will make is that the 
data are consistent with a trend in which the 
%AMG: why average or mean?
%average or 
%NL: We mention we try with the geometric mean as well (just below) and find 
%the same results.  No real formal reason against the median, except personally I think the median 
%is only particularly useful if you think you know something about what 
%the underlying (i.e. poorly sampled) distribution *should* be.  For example, 
%an asymmetric bi-modal distribution would pose a problem.  Do you think we need to get 
%into all that here, or just leave it up to a reviewer to bring up 
%if they think it is important?  I've been thinking about this issue a lot, so can discuss it...
%AMG: Especially if you've been thinking about this, let's discuss it briefly.  
mean cross-section for 
stellar encounters increases with increasing multiplicity.  
As with triples, this reduces the average time between 
encounters, and contributes to increasing the dynamical significance of 
higher-order multiplicity.  To show this trend, we calculate for each multiplicity the 
gravitationally-focused 
cross-section or, more specifically, the product of the total system mass and the geometric 
cross-section.  All calculations performed in this section 
correspond to the arithmetic mean, however we confirm our results 
using the geometric mean as well.  All uncertainties correspond 
to one standard 
%AMG: changed this, but be sure this is correct
%deviation 
error
of the mean.\footnote{We note that some of the distributions in our samples, and particularly those of the semi-major 
axes, are not symmetric about the mean.  Therefore their uncertainties 
about the mean are asymmetric.  However, 
the sample sizes are too small to properly calculate asymmetric error bars, and therefore we simply provide the standard 
errors of the means as our best estimates of the uncertainties.}
%AMG: edited the footnote.  See if this is OK with you.

First, as the number of stars increases, the \textit{geometric} cross-section 
(i.e. the diameter of the object) increases, and therefore the encounter 
time decreases.  We inspect the MSS sample for the star-forming 
association Taurus-Auriga \citep{kraus11}, and compute 
means and standard deviations of the mean for the maximum orbital separations (i.e. within a given 
system) for all binaries, triples, and quadruples.  These are 
$(1.3 \pm 0.4) \times 10^2$ AU, $(1.8 \pm 0.5) \times 10^3$ AU, and 
$(1.4 \pm 0.3) \times 10^3$ AU, 
respectively.  There are also two additional multiple star systems in this 
sample, namely a quintuple and a sextuple, with maximum orbital separations of 
3.1 $\times 10^3$ AU, and 3.8 $\times 10^3$ AU, respectively.  A similar analysis 
of the Multiple Star Catalogue \citep[MSC;][]{tokovinin97} yields mean 
maximum (i.e. highest-order) orbital separations for triples, quadruples, 
quintuples, sextuples, and septuples of $(1.8 \pm 0.1) \times 10^3$ AU, 
$(2.3 \pm 0.5) \times 10^3$ AU, $(9.0 \pm 1.7) \times 10^3$ AU, 
$(1.1 \pm 0.2) \times 10^4$ AU, and $(9.4 \pm 3.9) \times 10^3$ AU, 
respectively.

%sextuples, and septuples of $(1.0 \pm 0.1) \times 10^5$ AU, 
%$(1.3 \pm 0.4) \times 10^5$ AU, $(5.1 \pm 1.2) \times 10^5$ AU, 
%$(6.3 \pm 2.2) \times 10^5$ AU, and $(5.4 \pm 2.3) \times 10^5$ AU,
%respectively.  

%EDIT FIGURE CAPTION AND TEXT REGARDING FIGURE 2...AND FIGURE 1 ABOVE!!!

Second, the total stellar mass tends to increase with increasing multiplicity, which 
increases the \textit{gravitationally-focused} 
cross section, and thereby further decreases the encounter time.  
The mean system masses for the Taurus-Auriga sample (derived using the component masses 
given in Table 5 of \citealt{kraus11}) for the binaries, triples, quadruples, quintuple, 
and sextuple, respectively, are $1.0 \pm 0.1$ M$_{\odot}$, $1.3 \pm 0.1$ M$_{\odot}$, 
$2.0 \pm 0.3$ M$_{\odot}$, 2.9 M$_{\odot}$, and 3.2 M$_{\odot}$.  
For comparison, in the MSC the mean system masses for triples, quadruples, quintuples, 
and sextuples, respectively, are $4.5 \pm 0.1$ M$_{\odot}$, 
$5.0 \pm 0.6$ M$_{\odot}$, $5.6 \pm 0.6$ M$_{\odot}$, and 
$13.1 \pm 1.9$ M$_{\odot}$ (excluding systems with only minimum mass 
measurements).\footnote{We note that our results remain
unchanged with or without excluding these systems.} 
There is also a single septuple in the MSC sample with firm 
mass estimates for all 
components, yielding a total system mass of 28 M$_{\odot}$.

The key conclusion from these results is that the \textit{gravitationally-focused} 
cross-section for encounters systematically increases with increasing multiplicity.  
Figure~\ref{fig:fig3} shows the product of the mean total 
system mass and the mean maximum orbital separation, which is directly proportional 
to the gravitationally-focused cross-section, as a function of multiplicity for 
the Galactic field (top inset) and Taurus-Auriga (middle inset).  For comparison, 
%AMG: slight edit (since they're not actually cross sections)
%we show as well the mean cross-sections calculated for 
we also show the results from
the hydrodynamical 
simulations of \citet{bate12} using the values provided in their Table 3.  
The Spearman rank correlation coefficients for these samples are 
1.0 in all cases.  Thus, 
our proxy for the mean gravitationally-focused cross-section increases 
\textit{monotonically} with increasing multiplicity 
%AMG edited:
%, and this is consistent 
in both 
%with our 
of these observed MSS samples, 
which is also consistent with the theoretical predictions of \citet{bate12}.  

To better quantify the 
statistical significance of these correlations, we also performed weighted 
lines of best-fit
%AMG: added for clarity
to the $\log(m_{sys}a_{out})$ values shown in Figure~\ref{fig:fig3}.
For the Galactic field, Taurus-Auriga, and \citet{bate12} 
samples we find slopes of, respectively, 0.47 $\pm$ 0.07, 0.63 $\pm$ 0.09, and 
0.40 $\pm$ 0.10.  These are all inconsistent with zero at greater than 
the 3$\sigma$ confidence level.  To confirm this, we also calculated 
chi-squared values by comparing the 
%AMG: for consistency
$\log(m_{sys}a_{out})$ results
%observed (logarithmic) product of 
%the mean system mass and the mean maximum orbital separation 
to a flat 
%or constant 
distribution at the respective mean for each sample.  In all 
three cases, we 
%are able to 
rule out 
%AMG: added
the hypothesis
that the data are consistent with 
a flat distribution at very high confidence ($>$99.99\%). 
%AMG: slight edit 
%Clearly, the data suggests that a correlation exists between multiplicity and the 
In summary, these data show that the gravitationally-focused cross-section 
increases with increasing multiplicity at a statistically significant level.

%AMG: moved this here
The gravitationally-focused cross-section typically dominates 
over the geometric cross-section, especially for single stars and in low-velocity
dispersion environments.
However, with high-order multiples, the orbital separations can be so large that
both cross-sections can be equally important.
As discussed above, the data for these high-order multiples are also consistent with the 
geometric cross-section increasing with increased multiplicity, particularly for the MSC 
sample, where the data extend from triples to a septuple.  We note that the binding 
energies corresponding to the outer-most orbits of the MSSs in our sample are typically below 
the hard-soft boundary in their host cluster.  Thus, we expect that most of these systems 
are indeed bound objects.

%AMG: added, see what you think, and feel free to edit
%As is clear from the uncertainties on Figure~\ref{fig:fig3} and in the values given in the text, 
%further observations and larger sample sizes would be highly beneficial for confirming the statistical 
%significance of this apparent trend.  
%Nonetheless, 

The combination of the results presented in this section and the dynamical significance of triple stars 
shown in Section~\ref{dynsig} 
suggests that high-order multiples may be very important for many stellar dynamical processes in star 
clusters that have generally been attributed only to binary stars.  We discuss the implications of 
our result below.

\section{Discussion} \label{discussion}

%AMG: removed this given the paragraph above.  OK?
%We have confirmed that triple stars are as dynamically active in open 
%clusters as are single and binary stars.  We have further shown that the 
%data are consistent with a more general trend in which the 
%gravitationally-focused cross-section for encounters increases with 
%increasing multiplicity.  Below, we 
%explore the significance of these results for the dynamics of 
%star clusters.

Multiplicity is more likely
to be preserved during dynamical interactions in low-velocity-dispersion
environments.  This is because the hard-soft boundary is located at
very long orbital periods in such star clusters, so that even wide high-order
multiples are typically classified as dynamically hard.
We have verified this using a suite of numerical
scattering experiments involving single, binary, and triple stars
(see \citet{leigh12} for the details of the experiments).  By
keeping the orbital energies fixed and varying the relative velocity
at infinity, we find that decreasing the total
encounter energy on average increases the fraction of encounter
outcomes in which high-order multiples are produced or preserved.
This higher likelihood to produce and/or preserve MSSs suggests that 
high-order multiples 
should be more common and act as more efficient heat sources 
in low-velocity-dispersion environments such as open clusters (OCs).
%AMG: removed this, because I think we'd need at least one more sentence discussing how the velocity dispersion doesn't change very much with age or something like that if we want to get into the age argument.  And it doesn't seem very central to the argument.  OK with you?
%, almost independent of their age.

%AMG: reorganized paragraph a bit
\citet{leigh12} recently showed 
that the probability that a direct collision will occur between 
any two stars during an encounter scales as $N^2$, where $N$ is 
the number of interacting stars.  %\footnote{Or, more 
%accurately, the probability likely scales as $N(N-1)$, which is derived from ``N choose 2''.  In 
%\citet{leigh12}, we were unable to distinguish between the scalings $N(N-1)$ 
%and $N^2$ at a statistically significant level.}  
% level due to the number of free parameters 
%needed in our fits to the data.}  
%AMG: is this new?  I'll trust you on this, but can you give me some background. I don't remember this part.
Therefore, on a per encounter basis, a higher multiplicity translates into an 
increased probability 
of direct stellar collisions, which may be observed as stellar exotica such as BSs.
%Moreover, resonant encounters are known to lead to stellar collisions, which 
%can produce various types of stellar exotica, 
%particularly BSs.  
Several peculiar examples of 
MSSs containing BSs have recently been identified in OCs.  
Their existence is difficult to explain without at least some 
involvement from dynamical interactions.  For example, the suspected 
triple system S1082 in the old OC M67 appears to contain 
two BSs \citep{vandenberg01,sandquist03}.  A triple system in which 
the outer companion is a BS was recently found in the OC 
NGC 6819 by \citet{talamantes10}.  A fascinating 
binary containing two BSs was also reported by \citet{mathieu09} in 
the old OC NGC 188.  In all of these cases, mass transfer 
alone could not have produced the systems containing these
BSs, whereas a collisional origin 
or at least a subsequent stellar encounter(s) 
involving high-order multiples may potentially account for their 
existence.  

In fact, the multiple-star population may play a large role in governing 
both the creation \textit{and destruction} rates of stellar exotica, 
in particular those that rely on mass transfer in compact binaries. 
The results of \citet{leigh12} also show that physical collisions during dynamical 
encounters, which become more likely with increasing multiplicity, offer a 
mechanism for destroying compact binaries.  This stems from the fact that, if 
a collision occurs during an encounter involving three or more 
stars, it is the most compact orbit going into the interaction that will
typically be destroyed 
\citep[e.g.][]{valtonen06}.  
The details of the balance between creation and destruction of compact 
binaries resulting from encounters with MSSs in clusters is an important
question that can be answered by the next generation of $N$-body models 
that include significant populations of high-order multiples \citep[e.g.][]{geller13}.

On a more global scale, future $N$-body models will also show how high-order MSSs
influence the overall dynamical evolution of star clusters.
For example,
$N$-body simulations have shown that the hardening of compact binaries through 
successive dynamical encounters in the dense cores of GCs 
can halt core collapse \citep[e.g.][]{hut83}.  This 
evolutionary phase is often called ``binary burning'' 
\citep[e.g.][]{fregeau09}, since these hardening interactions act to prevent 
any further increase in the central cluster density.  
However, the collisional cross-section for a compact binary is very small,
so that the time-scale for it to undergo a direct dynamical interaction
with a single star can be very long.  Consequently, the onset of the binary 
burning phase is expected to exceed a Hubble time for most of the GCs in the Milky Way 
\citep{fregeau09}.  However, if most 
compact binaries are members of high-order multiples, then the time-scale
for them to undergo direct encounters can be quite short due to the much
larger cross-section of their parent MSS.  
The wide outer orbit of a triple effectively acts as a ``net'', drawing 
stars in where they can be scattered and interact 
resonantly with the close inner binary of the triple \citep{leigh11,moeckel13}.  In this 
way, the binding energy of the compact inner binary of the triple can be tapped, and 
re-distributed throughout the cluster via relaxation processes (once the higher-order 
resonantly interacting multiple system has decayed into stable components).  
The role played by high-order multiples as heat sources in clusters 
should be the most 
pronounced in low-velocity-dispersion environments, like OCs.  
As discussed above, in such clusters even very wide orbits, which are a 
prerequisite for dynamical stability within high-order multiples, are often 
classified as dynamically hard.  Our results suggest that triples 
may undergo dynamical interactions roughly as frequently as binary stars, and 
hence could be relevant for the overall evolution of certain clusters.  
%However, we stress that our results do not show 
%that triples are necessarily the dominant heat source in clusters, only that 
%they should be undergoing dynamical interactions roughly as frequently as 
%binary stars, and hence could be relevant for the overall cluster evolution in 
%certain cluster 
%environments.  
Future detailed $N$-body modeling will be necessary in order to quantify the 
role of triples as heat sources and their significance for 
the dynamical evolution of star clusters.

\section{Summary} \label{summary}

In this paper, we address the question:  
How important are high-order MSSs to the dynamics of star clusters and the 
production/destruction of compact binaries and stellar exotica?  
The key conclusion 
is that dynamical encounters 
involving triple stars should be roughly as common as encounters involving only 
single and binary stars in low- to moderate-density star 
clusters.  
Furthermore, the available data on higher-order multiples 
are consistent with a trend in which 
the gravitationally-focused cross-section for encounters increases with 
increasing multiplicity.  These results suggest 
that simulations of star cluster evolution should include 
high-order multiples 
in the initial conditions 
%NL: I removed the below remark in parantheses, since I know it will piss off Pavel Kroupa.
%(perhaps consistent with Taurus-Auriga) 
and allow such systems 
to evolve dynamically, and be created and destroyed, throughout the cluster evolution.
This will be a crucial next 
step in the development of realistic star cluster simulations.

%\vspace*{1 mm}

\section*{Acknowledgments}

We would like to thank both Kaitlin Kratter and Adam Kraus for valued input 
and advice throughout the preparation of this manuscript, as well as David 
Latham for useful discussions.  
%AMG: removed this--was Kaitlin, right?
%Support for this work was provided by NASA through Hubble Fellowship 
%grant \#HST-HF-51306.01 awarded by the Space Telescope Science Institute, 
%which is operated by the Association of Universities for Research in 
%Astronomy, Inc., for NASA, under contract NAS 5- 26555.

%\chapterbib
%\vspace*{1 mm}

\bsp

\label{lastpage}

\end{document}